\documentclass[toc]{PoS}
\usepackage{cite,amsmath}

\newcommand{\as}{\alpha_{\rm s}}
\def\MSbar{\overline{\mathrm{MS}}}
\def\ep{\epsilon}

\def\ca{{C^{}_A}}
\def\cf{{C^{}_F}}
\def\tf{{T^{}_F}}
\def\nf{{n^{}_{\! f}}}
\def\nl{{n^{}_{\! l}}}
\def\nh{{n^{}_{\! h}}}

\title{Massive QCD amplitudes at higher orders}

\ShortTitle{Massive QCD amplitudes at higher orders}

\author{\speaker{S. Moch}\\
        Deutsches Elektronen--Synchrotron, DESY   \\
        Platanenallee 6, D--15738 Zeuthen, Germany\\
        E-mail: \email{sven-olaf.moch@desy.de}
}

\author{A. Mitov \\
        Department of Mathematical Sciences, University of Liverpool \\ 
        Liverpool L69 3BX, United Kingdom \\
        E-mail: \email{Alexander.Mitov@liverpool.ac.uk}
}

\abstract{
We consider the factorization properties of on-shell QCD amplitudes
with massive partons in the limit when all kinematical invariants
are large compared to the parton mass and discuss the structure of
their infrared singularities. The dimensionally regulated soft poles
and the large collinear logarithms of the parton masses
exponentiate to all orders. Based on this factorization a simple
relation between massless and massive scattering amplitudes in gauge
theories can be established. We present recent applications of this
relation for the calculation of the two-loop virtual QCD corrections
to the hadro-production of heavy quarks.
}

\FullConference{
``Matter To The Deepest'', September 5-11 , 2007, Ustron, Poland.\\
8th International Symposium on Radiative Corrections (Radcor 2007), October 1-5, 2007, Florence, Italy.}

\begin{document}

%
%
\section{Introduction}
\label{sec:introduction}

Amplitudes for hard scattering processes in Quantum Chromodynamics
(QCD) are of basic importance both for theory and phenomenology and
precision predictions for them must include higher-order quantum
corrections. An important aspect in explicit computations are the
singular limits of amplitudes at higher loops. Here, one has to
consider two types of limits, soft and collinear, related to the
emission of gluons with vanishing energy and to collinear parton
radiation off massless hard partons, respectively.

For massless QCD amplitudes the corresponding singularities are
regularized by working in $d$ dimensions and appear as explicit
poles in $(d-4)$. Typically two powers in $1/(d-4)$ are generated
per loop. When massive particles are involved, some of the collinear
singularities are screened by the parton masses, which gives rise to
large logarithmically enhanced contributions. In both cases the
structure of the singularities can be understood from the
factorization property of QCD and it can be predicted to all orders
based on a small number of perturbatively calculable anomalous
dimensions. Moreover, factorization gives rise to an extremely
simple universal multiplicative relation between a massless
amplitude and its massive version in the limit when the parton
masses are small with respect to all other kinematical invariants.
This relation can be employed to derive virtual QCD corrections at
higher loops including all logarithms in the heavy quark mass as
well as all constant (mass-independent) contributions. We
demonstrate the predictive power of factorization with recent
results for the two-loop QCD amplitudes for heavy-quark production
in hadronic collisions.

%
%
\section{Factorization of QCD amplitudes}
\label{sec:factorization}

We are interested in general $2 \to n$ scattering processes of partons $p_i$
\begin{equation}
\label{eq:QCDscattering}
{\rm p}: \qquad
p_1 + p_2 \:\:\rightarrow\:\: p_3 + \dots + p_{n+2}
\, .
\end{equation}
The corresponding scattering amplitude ${\cal M}_{\rm p}$ depends on the
set of fixed external momenta $\{ k_i \}$, masses $\{ m_i \}$ and color quantum numbers $\{ c_i \}$,
\begin{eqnarray}
\label{eq:QCDamplitude}
| {\cal M}_{\rm p} \rangle
&\equiv&
{\cal M}_{\rm p}\left(\{ k_i \},\{ m_i \},{\{c_i\}},{Q^2 \over \mu^2},\as(\mu^2),\ep \right)
\, ,
\end{eqnarray}
as well as on the strong coupling constant $\as$, the renormalization scale $\mu$
and the parameter $\ep$ of dimensional regularization, $d=4-2 \ep$.
Also, we denote explicitly the hard scale $Q$ of the process
typically related to the center-of-mass energy, e.g. $Q = \sqrt{s}$ with
$s = (k_1 + k_2)^2$.

Let us briefly recall the factorization of on-shell amplitudes
for {\it massless} partonic processes~\cite{Catani:1998bh,Sterman:2002qn}.
In $d$-dimensions we can write Eq.~(\ref{eq:QCDamplitude}) as a product of functions
${\cal J}_{\rm p}^{(m=0)}$,
${\cal S}_{\rm p}^{(m=0)}$
and ${\cal H}^{\rm[p]}$,
\begin{eqnarray}
\label{eq:QCDfacamplitude-zero}
| {\cal M}_{\rm p}\rangle^{(m=0)} \, = \, 
{\cal J}_{\rm p}^{(m=0)}\left({Q^2 \over \mu^2},\as(\mu^2),\ep \right)
{\cal S}_{\rm p}^{(m=0)}\left(\{ k_i \},{Q^2 \over \mu^2},\as(\mu^2),\ep \right)
| {\cal H}_{\rm p} \rangle
\, ,
\end{eqnarray}
where we use matrix notation suppressing the color indices.
The jet function ${\cal J}_{\rm p}^{(m=0)}$ depends only on the external
partons. It collects all collinearly sensitive contributions and is color-diagonal.
Coherent soft radiation arising from the overall color flow is summarized by
the soft function ${\cal S}_{\rm p}^{(m=0)}$, which is a matrix in color space.
The short-distance dynamics of the hard scattering
is described by the (infrared finite) hard function ${\cal H}_{\rm p}$,
which to leading order is simply proportional to the Born amplitude.

The factorization formula~(\ref{eq:QCDfacamplitude-zero})
organizes the singularity structure of any {\it massless} QCD amplitude.
From the operator definitions for the functions ${\cal J}_{\rm p}^{(m=0)}$ and ${\cal S}_{\rm p}^{(m=0)}$
and the corresponding renormalization group properties one derives evolution equations.
The solution of the latter gives rise to an all-order exponentiation
in terms of well-known anomalous dimensions.
As an upshot, all $1/\epsilon$ terms related to the emission of gluons with vanishing energy
and to collinear parton radiation off massless hard partons, respectively,
exponentiate to all orders in perturbation theory, see e.g.~\cite{Sterman:2002qn,MertAybat:2006mz}.

Most important for our considerations is the jet function, which
contains all collinear contributions from the external partons. It
is therefore of the form
\begin{equation}
  \label{eq:jetfactor}
  {\cal J}_{\rm p}^{(m=0)}
  \, = \,
  \prod_{i\in\ \{{\rm all}\ {\rm legs}\}}\,
  {\cal J}_{[i]}^{(m=0)}
  \, = \,
  \prod_{i\in\ \{{\rm all}\ {\rm legs}\}}\,
  \left(
    {\cal F}_{[i]}^{(m=0)}
  \right)^{1 \over 2}
  \, ,
\end{equation}
where $i=q,g$ for quarks and gluons.
${\cal J}_{[i]}^{(m=0)}$ is the individual jet function of each external parton,
which by definition, one identifies with the (gauge invariant) form factor of
a quark or a gluon, ${\cal F}_{[i]}^{(m=0)}$.
Of course, the latter function is well-known in QCD, see e.g.~\cite{Moch:2005id,Moch:2005tm}.

When masses are introduced the picture described above gets
modified. However, the basic factorization of the QCD amplitude in
jet, soft and hard function from Eq.~(\ref{eq:QCDfacamplitude-zero})
can be retained. With the exception of contributions related to
heavy quark loops (see below) in the presence of a hard scale $Q$ we
can write for the partonic process~(\ref{eq:QCDscattering})
\begin{eqnarray}
\label{eq:QCDfacamplitude}
| {\cal M}_{\rm p}\rangle^{(m)} \, = \,
{\cal J}_{\rm p}^{(m)}\left({Q^2 \over \mu^2},\{ m_i \},\as(\mu^2),\ep \right)
{\cal S}_{\rm p}^{(m)}\left(\{ k_i \},{Q^2 \over \mu^2},\as(\mu^2),\ep \right)
| {\cal H}_{\rm p} \rangle
\, ,
\end{eqnarray}
where all non-trivial mass dependence enters in the functions
${\cal J}_{\rm p}^{(m)}$ and ${\cal S}_{\rm p}^{(m)}$, while
power suppressed terms in the parton masses are neglected in ${\cal H}_{\rm p}$.
The jet function for massive partons can be defined
in complete analogy to Eq.~(\ref{eq:jetfactor}), i.e. we identify
${\cal J}_{[i]}^{(m)}$ with the massive form factor ${\cal F}_{[i]}^{(m)}$.
This guarantees exponentiation with largely the
same anomalous dimensions as in the massless case.
Also the soft anomalous dimensions which govern
the soft function have a smooth limit for vanishing parton masses.

In summary, from comparison of Eqs.~(\ref{eq:QCDfacamplitude-zero})
and~(\ref{eq:QCDfacamplitude}) one can deduce a remarkably simple
relation between a massless and a massive amplitude in the
small-mass limit. Thus, QCD factorization provides us with
\cite{Mitov:2006xs}
\begin{eqnarray}
\label{eq:Mm-M0}
{\cal M}_{\rm p}^{(m)} &=&
\prod_{i\in\ \{{\rm all}\ {\rm legs}\}}\,
  \left(
    Z^{(m\vert0)}_{[i]}
  \right)^{1 \over 2}\,
  \times\
{\cal M}_{\rm p}^{(m=0)}\, ,
\end{eqnarray}
where we have again suppressed the color indices.
${\cal M}_{\rm p}^{(m=0)}$ and the corresponding massive amplitude
${\cal M}_{\rm p}^{(m)}$ in the small mass limit
$m^2 \ll Q^2$ are multiplicatively related by a universal process independent function $Z^{(m\vert 0)}_{[i]}$.
With the definitions for the jet functions in Eq.~(\ref{eq:jetfactor}),
it is directly given in terms of the respective form factors,
\begin{equation}
\label{eq:Z}
Z^{(m\vert0)}_{[i]}\left({m^2 \over \mu^2},\as,\ep \right)
\, = \,
{\cal F}_{[i]}^{(m)}\left({Q^2\over \mu^2},{m^2\over\mu^2},\as,\ep \right)
\left({\cal F}_{[i]}^{(m=0)}\left({Q^2\over \mu^2},\as,\ep \right)\right)^{-1}
\, ,
\end{equation}
where the index $i$ denotes the (massive) parton and $\as = \as(\mu^2)$.
The process-independence is manifest in Eq.~(\ref{eq:Z}),
because $Z^{(m\vert0)}_{[i]}$ is only a function of the (process-independent) ratio of scales $\mu^2/m^2$.
The (process-dependent) scale $Q$ cancels completely between the massive and the massless form factors.

Eq.~(\ref{eq:Mm-M0}) can be used to predict any massive amplitude
from the known massless one, which is a great advantage in practice,
as the latter is much easier to compute. Moreover,
Eq.~(\ref{eq:Mm-M0}) includes not only the singular terms and the
logarithms in the massive amplitude but extends even to the
mass-independent constant contributions. However, there is one
important side condition on Eqs.~(\ref{eq:Mm-M0}) and (\ref{eq:Z}),
concerning all terms proportional to the number of heavy quarks
$\nh$. These two-loop contributions are excluded explicitly from the
definition of $Z^{(m\vert 0)}_{[i]}$. In order to incorporate them
additional process dependent terms appear as has been shown e.g. in
QED for Bhabha scattering~\cite{Becher:2007cu}.

%
%
\section{Hadro-production of heavy quarks}
\label{sec:hadro-production}

As an application of the formalism developed, we consider
the pair-production of heavy quarks in the $q {\bar q}$-annihilation
and the gluon fusion channel,
\begin{eqnarray}
\label{eq:ppQQ}
{\rm q}: \qquad
q(k_1) + {\bar q}(k_2) &\rightarrow & Q(k_3,m) + {\bar Q}(k_4,m) \, ,
\\
{\rm g}: \qquad
g(k_1) + g(k_2) &\rightarrow & Q(k_3,m) + {\bar Q}(k_4,m) \, ,
\nonumber
\end{eqnarray}
where $k_i$ denote the on-shell parton momenta and $m$ the mass of
the heavy quark, thus $k_1^2 = k_2^2 = 0$ and $k_3^2 = k_4^2 = m^2$.
Energy-momentum conservation implies $k_1^\mu + k_2^\mu = k_3^\mu +
k_4^\mu$ and we consider the scattering amplitude for the
processes~(\ref{eq:ppQQ}) in QCD perturbation theory,
\begin{eqnarray}
  \label{eq:Mexp}
  | {\cal M}_{\rm p} \rangle^{(m)}
  \, = \,
  4 \pi \as \biggl[
  | {\cal M}_{\rm p}^{(0)} \rangle^{(m)}
  + \biggl( {\as \over 2 \pi} \biggr) | {\cal M}_{\rm p}^{(1)} \rangle^{(m)}
  + \biggl( {\as \over 2 \pi} \biggr)^2 | {\cal M}_{\rm p}^{(2)} \rangle^{(m)}
  + {\cal O}(\as^3)
  \biggr]
\, ,
\end{eqnarray}
which defines the series expansion in the strong coupling $\as =
\as(\mu^2)$ and $\mu$ is the renormalization scale. As usual the
$\MSbar$-scheme for the coupling constant renormalization is
employed and the mass $m$ is taken to be the pole mass.

It is convenient to define the function ${\cal A}_{\rm p}(\epsilon, m, s, t, \mu)$
for the squared amplitudes summed over spins and colors as
\begin{eqnarray}
\label{eq:Msqrd}
\overline{\sum |{\cal M}_{\rm p}|^2}
&=&
{\cal A}_{\rm p}(\epsilon, m, s, t, \mu)\, ,\qquad\qquad {\rm p} = {\rm q,g}
\, .
\end{eqnarray}
${\cal A}_{\rm p}$ is a function of the Mandelstam variables $s$, $t$ and $u$ given by
\begin{equation}
\label{eq:Mandelstam}
s = (p_1+p_2)^2\, , \qquad
t  = (p_1-p_3)^2 - m^2\, , \qquad
u  = (p_1-p_4)^2 - m^2\, ,
\end{equation}
and has a perturbative expansion similar to Eq.~(\ref{eq:Mexp}),
\begin{eqnarray}
\label{eq:Aexp}
{\cal A}_{\rm p}^{(m)}(\epsilon, m, s, t, \mu)
  \, = \,
16 \pi^2 \as^2
\left[
  {\cal A}_{\rm p}^{4,(m)}
  + \biggl( {\as \over 2 \pi} \biggr) {\cal A}_{\rm p}^{6,(m)}
  + \biggl( {\as \over 2 \pi} \biggr)^2 {\cal A}_{\rm p}^{8,(m)}
  + {\cal O}(\as^{3})
\right]
\, .
\end{eqnarray}
In terms of the amplitudes the expansion coefficients in Eq.~(\ref{eq:Aexp})
may be expressed as
\begin{eqnarray}
\label{eq:A4def}
{\cal A}_{\rm p}^{4,(m)} &=&
\langle {\cal M}_{\rm p}^{(0)} | {\cal M}_{\rm p}^{(0)} \rangle^{(m)}
\, , \\
\label{eq:A6def}
{\cal A}_{\rm p}^{6,(m)} &=&
\langle {\cal M}_{\rm p}^{(0)} | {\cal M}_{\rm p}^{(1)} \rangle^{(m)}
+ \langle {\cal M}_{\rm p}^{(1)} | {\cal M}_{\rm p}^{(0)} \rangle^{(m)}
\, , \\
\label{eq:A8def}
{\cal A}_{\rm p}^{8,(m)} &=&
\langle {\cal M}_{\rm p}^{(1)} | {\cal M}_{\rm p}^{(1)} \rangle^{(m)}
+ \langle {\cal M}_{\rm p}^{(0)} | {\cal M}_{\rm p}^{(2)} \rangle^{(m)}
+ \langle {\cal M}_{\rm p}^{(2)} | {\cal M}_{\rm p}^{(0)} \rangle^{(m)}
\, ,
\end{eqnarray}
where the results for ${\cal A}_{\rm p}^{6,(m)}$ have been presented e.g. in~\cite{Korner:2002hy,Bernreuther:2004jv} and
the so-called loop-by-loop contribution in ${\cal A}^{8,(m)}$ can be found in~\cite{Korner:2005rg},
both results with the complete dependence on the heavy-quark mass.
The new contribution is the real part of $\langle {\cal M}_{\rm p}^{(0)} | {\cal M}_{\rm p}^{(2)} \rangle^{(m)}$
up to powers ${\cal O}(m)$ in the heavy-quark mass~\cite{Czakon:2007ej,Czakon:2007wk}.

In order to obtain $\langle {\cal M}_{\rm p}^{(0)} | {\cal M}_{\rm p}^{(2)}
\rangle^{(m)}$ from Eq.~(\ref{eq:Mm-M0}), we have to construct the appropriate
functions $Z^{(m\vert 0)}_{[i]}$ from the on-shell heavy-quark form factor
and the massless on-shell ones,
all results being known~\cite{Bernreuther:2004ih,Moch:2005id,Moch:2005tm,Czakon:2007wk}
to sufficient orders in $\as$ and powers of $\epsilon$.
An explicit expression for
\begin{equation}
\label{eq:ZQ}
Z^{(m\vert0)}_{[Q]} \, = \, 1 + {\as \over 2 \pi} \,
Z^{(1)}_{[Q]} + \left( {\as \over 2 \pi} \right)^2\, Z^{(2)}_{[Q]}
\, +{\cal O}(\as^3)\, ,
\end{equation}
up to two loops is known~\cite{Mitov:2006xs}.
As mentioned above, the definition~(\ref{eq:ZQ}) accounts in particular for all fermionic terms
except for those linear in $\nh$.
The leading $\nf$ terms $\sim(\nf\as)^n$ for the process $gg\to Q{\bar Q}$ in Eq.~(\ref{eq:ppQQ})
can also be predicted,
where we denote the total number of flavors with $\nf$,
which is the sum of $\nl$ light and $\nh$ heavy quarks.
Keeping only terms quadratic in $\nh$ and/or $\nf = \nh+\nl$ one has up to two loops,
\begin{equation}
\label{eq:Zg}
Z^{(m\vert0)}_{[g]} \, = \, 1 + {\as \over 2 \pi} \,
Z^{(1)}_{[g]} + \left( {\as \over 2 \pi} \right)^2\, Z^{(2)}_{[g]}
\, +{\cal O}(\as^3)\, ,
\end{equation}
where
\begin{equation}
\label{eq:Zgtwo}
Z^{(2)}_{[g]}\, =\, \left( Z^{(1)}_{[g]} \right)^2
+\, {2\over 3\epsilon}\,\nf\tf\, Z^{(1)}_{[g]} + {\cal O}(\nh^1 \times \nl^0)\, ,
\end{equation}
with $Z^{(1)}_{[g]} \sim \nh$ known from~\cite{Mitov:2006xs,Czakon:2007wk}.
Note that $Z^{(1)}_{[g]}$ is also equal to the ${\cal O}(\as)$ term in the gluon wave
function renormalization constant $Z_3$.

Exploiting the predictive power of the relation Eq.~(\ref{eq:Mm-M0}) and applying
it to the processes~(\ref{eq:ppQQ}) we get
\begin{eqnarray}
\label{eq:A8mtoA80q}
{\lefteqn{
2 {\rm Re}\, \langle {\cal M}_{\rm q}^{(0)} | {\cal M}_{\rm q}^{(2)} \rangle^{(m)}
\, = \,
2 {\rm Re}\, \langle {\cal M}_{\rm q}^{(0)} | {\cal M}_{\rm q}^{(2)} \rangle^{(m=0)}
}} \\
& & + Z_{[Q]}^{(1)} {\cal A}_{\rm q}^{6,(m=0)}\ +\ 2 Z_{[Q]}^{(2)} {\cal A}_{\rm q}^{4,(m=0)}
+\ {\cal O}(\nh^1 \times \nl^0) +\ {\cal O}(m) \, ,
\nonumber\\
\label{eq:A8mtoA80g}
{\lefteqn{
2 {\rm Re}\, \langle {\cal M}_{\rm g}^{(0)} | {\cal M}_{\rm g}^{(2)} \rangle^{(m)}
\, = \,
2 {\rm Re}\, \langle {\cal M}_{\rm g}^{(0)} | {\cal M}_{\rm g}^{(2)} \rangle^{(m=0)}
}} \\
& &
+ \left(Z^{(1)}_{[Q]} + Z^{(1)}_{[g]}\right) {\cal A}_{\rm g}^{6,(m=0)}
+ 2 \left(Z^{(2)}_{[Q]} + Z^{(2)}_{[g]} + Z^{(1)}_{[Q]} Z^{(1)}_{[g]}\right) {\cal A}_{\rm g}^{4,(m=0)}\
+\ {\cal O}(\nh^1 \times \nl^0) +\ {\cal O}(m)
\, ,
\nonumber
\end{eqnarray}
which assumes the hierarchy of scales $m^2 \ll s,t,u$,
i.e. we neglect terms ${\cal O}(m)$.
Eqs.~(\ref{eq:A8mtoA80q}) and (\ref{eq:A8mtoA80g}) predict the complete real part of the squared amplitudes
$\langle {\cal M}_{\rm p}^{(0)} | {\cal M}_{\rm p}^{(2)} \rangle^{(m)}$
except (as indicated) for those terms, which are linear in $\nh$.
The real part of the two-loop massless amplitudes
$\langle {\cal M}_{\rm p}^{(0)} | {\cal M}_{\rm p}^{(2)} \rangle^{(m=0)}$
are computed in~\cite{Anastasiou:2000kg,Anastasiou:2001sv}.
The finite remainders of the latter agree with the corresponding terms constructed from the two-loop helicity
amplitudes calculated of~\cite{Bern:2003ck,DeFreitas:2004tk} after the infrared subtraction procedure
is performed.

In order to arrive at a complete prediction for $\langle {\cal
M}_{\rm p}^{(0)} | {\cal M}_{\rm p}^{(2)} \rangle^{(m)}$ including
all heavy-quark loop corrections, the factorization approach in
Eqs.~(\ref{eq:A8mtoA80q}), (\ref{eq:A8mtoA80g}) is supplemented by a
direct calculation of all necessary massive Feynman diagrams as an
expansion in the small mass. The advantage of this approach is an
independent check of Eqs.~(\ref{eq:A8mtoA80q}) and
(\ref{eq:A8mtoA80g}) as well as of the corresponding massless
results. It relies on the reduction of integrals to a set of masters
with the Laporta algorithm \cite{Laporta:2001dd}, the subsequent
construction of Mellin-Barnes representations for all the integrals,
see
e.g.~\cite{Smirnov:1999gc,Tausk:1999vh,Czakon:2004wm,Czakon:2005rk,Czakon:2006pa}
and the summation of series representations~\cite{Moch:2005uc} or
the application of the PSLQ algorithm~\cite{pslq:1992}.

We are now able to give the result for the
interference of the two-loop and Born amplitude
for the scattering processes~(\ref{eq:ppQQ}).
For a ${\rm{SU}}(N)$-gauge theory with $N$ denoting the number of colors,
one has $\ca = N$, $\cf = (N^2-1)/2\*N$ and $\tf = 1/2$ and, 
as mentioned above, the total number of flavors $\nf = \nl + \nh$ is the sum of
$\nl$ light and $\nh$ heavy quarks.
Thus, for $q{\bar q} \to Q {\bar Q}$ and $gg \to Q {\bar Q}$ we have
\begin{eqnarray}
  \label{eq:ReM0x2q}
  2 {\rm Re}\, \langle {\cal M}_{\rm q}^{(0)} | {\cal M}_{\rm q}^{(2)} \rangle^{(m)} 
  & = &
      2 (N^2-1) \biggl(
      N^2 A_{\rm q} + B_{\rm q}  + {1 \over N^2} C_{\rm q}
  + N \nl D_{{\rm q},l} + N \nh D_{{\rm q},h}
  \\
  & &
  + {\nl \over N} E_{{\rm q},l} + {\nh \over N} E_{{\rm q},h} + (\nl+ \nh)^2 F_{\rm q}
  \biggr)
  \, ,
\nonumber\\
  \label{eq:ReM0x2g}
  2 {\rm Re}\, \langle {\cal M}_{\rm g}^{(0)} | {\cal M}_{\rm g}^{(2)} \rangle^{(m)} 
  & = &
(N^2-1) \biggl(
N^3 A_{\rm g} + N B_{\rm g}  + {1 \over N} C_{\rm g} + {1 \over N^3} D_{\rm g}
+ N^2 \nl E_{{\rm g},l} + N^2 \nh E_{{\rm g},h}
\\
& &
+ \nl F_{{\rm g},l} + \nh F_{{\rm g},h}
+ {\nl \over N^2} G_{{\rm g},l} + {\nh \over N^2} G_{{\rm g},h}
\nonumber\\
& &
+ N \nl^2 H_{{\rm g},l} + N \nl \nh\, H_{{\rm g},lh} + N \nh^2 H_{{\rm g},h}
+ {\nl^2 \over N} I_{{\rm g},l} + {\nl \nh \over N} I_{{\rm g},lh} + {\nh^2 \over N} I_{{\rm g},h}
\biggr)
\, ,
\nonumber
\end{eqnarray}
and all explicit expressions can be found in~\cite{Czakon:2007ej,Czakon:2007wk}.

As explained above, the factorization approach provides results for
all coefficients except the terms linear in $\nh$. These are
$D_{{\rm q},h}$ and $E_{{\rm q},h}$ in Eq.~(\ref{eq:ReM0x2q}) and
$E_{{\rm g},h}$, $F_{{\rm g},h}$ and $G_{{\rm g},h}$ in
Eq.~(\ref{eq:ReM0x2g}), which have been obtained from a direct
calculation of the massive loop integrals as briefly sketched above.
Furthermore, to have an independent cross check of the factorization
formulae~(\ref{eq:A8mtoA80q}) and (\ref{eq:A8mtoA80g}) the direct
computation of Feynman diagrams has also been extended to the
coefficients $A_{\rm q}$, $D_{{\rm q},l}$, $E_{{\rm q},l}$ and
$F_{\rm q}$ in Eq.~(\ref{eq:ReM0x2q}) and to $A_{\rm g}$, $E_{{\rm
g},l}$, $H_{{\rm g},l}$, $H_{{\rm g},lh}$, $H_{{\rm g},h}$, $I_{{\rm
g},l}$, $I_{{\rm g},lh}$, and $I_{{\rm g},h}$ in
Eq.~(\ref{eq:ReM0x2g}). Of course, for the coefficients tested we
have found full agreement between both methods.

%
%
\section{Conclusions}
\label{sec:conclusions}

We have presented a discussion of the singular behavior of on-shell
QCD amplitudes with massive particles at higher orders and we have
emphasized the strong similarities between scattering amplitudes
with massless and massive partons in the small mass limit. In this
regime, factorization relates the two amplitudes multiplicatively by
a universal process independent function $Z^{(m\vert 0)}_{[i]}$. The
present results for amplitudes generalize the massless formulae known
previously~\cite{Catani:1998bh,Sterman:2002qn} and they extend
one-loop massive results~\cite{Catani:2000ef} to all
orders. For cross sections the analogous property~\cite{Mele:1990cw} 
is now known through next-to-next-to-leading order~\cite{Melnikov:2004bm,Mitov:2004du}.

We have illustrated the predictive power of the factorization ansatz
with new results for heavy-quark hadro-production at two-loops in
QCD~\cite{Czakon:2007ej,Czakon:2007wk}. The results include all
logarithms in the mass $m$ as well as the constant terms and they
can be used as a strong check of any future complete calculation.
Moreover, if combined with the threshold behavior of the amplitude,
which is not known at present they can serve as a well founded basis
for quantitative predictions for, say, top production at LHC to
next-to-next-to-leading order. In order to obtain physical cross
sections, the virtual amplitudes considered here have to be combined
with the corresponding real emission contributions, of course.

In a similar spirit, one can also derive higher order QCD corrections to new heavy (colored)
particles, like squark or gluino production in supersymmetric extensions of
the Standard Model.
Also the approach taken here may prove useful in the future in formulating
subtraction schemes with massive partons for the real emission contributions
beyond one loop.

%
%

\providecommand{\href}[2]{#2}\begingroup\raggedright\endgroup

\end{document}